\documentstyle[12pt]{article}

\newcommand{\eq}{\begin{equation}}
\newcommand{\en}{\end{equation}}
\newcommand{\bea}{\begin{eqnarray}}
\newcommand{\eea}{\end{eqnarray}}
\newcommand{\spz}{\hspace{0.7cm}}
\newcommand{\virg}{\spz,\spz}

\newcommand{\half}{{\textstyle\frac{1}{2}}}

\newcommand{\mn}{{\cal M}A^{(+)}}
\newcommand{\prf}{H^{(\pi)}}

\newcommand{\bZ}{\mbox{\bf Z}}
\newcommand{\buno}{\mbox{\bf 1}}

\newcommand{\cA}{{\cal A}}

\newcommand{\cM}{{\cal M}}

\newcommand{\cV}{{\cal V}}

\newcommand{\NP}[1]{Nucl.\ Phys.\ {\bf #1}}
\newcommand{\PL}[1]{Phys.\ Lett.\ {\bf #1}}

\newcommand{\CMP}[1]{Comm.\ Math.\ Phys.\ {\bf #1}}

\newcommand{\IJMP}[1]{Int.\ J.\ Mod.\ Phys.\ {\bf #1}}

\begin{document}
\sloppy
\renewcommand{\thefootnote}{\fnsymbol{footnote}}

\newpage
\setcounter{page}{1}

\vspace{0.7cm}
\begin{flushright}
DFUB 95-09\\
November 1995
\end{flushright}
\vspace*{1cm}
\begin{center}
{\bf INFRARED BEHAVIOUR OF MASSLESS INTEGRABLE FLOWS ENTERING THE
MINIMAL MODELS FROM $\phi_{31}$}\\
\vspace{1.8cm}
{\large G.\ Feverati, E.\ Quattrini\footnote{Present address (until
April 16, 1996): Physikalisches Institut der Universit\"at Bonn,
Nu{\ss}allee 12, D-5300 Bonn, Germany\\
E-mail: feverati@bo.infn.it, quattrini@bo.infn.it, ravanini@bo.infn.it}
and F.\ Ravanini}\\
\vspace{.5cm}
{\em Sezione I.N.F.N. e Dip. di Fisica - Univ. di Bologna\\
     Via Irnerio 46, I-40126 BOLOGNA, Italy}
\end{center}
\vspace{1cm}

\renewcommand{\thefootnote}{\arabic{footnote}}
\setcounter{footnote}{0}
 
\begin{abstract}
It is known that any minimal model $M_p$ receives along its
$\phi_{31}$ irrelevant direction {\em two} massless integrable flows:
one from $M_{p+1}$ perturbed by $\phi_{13}$, the other from $Z_{p-1}$
parafermionic model perturbed by its generating parafermion field. By
comparing Thermodynamic Bethe Ansatz data and ``predictions'' of
infrared Conformal Perturbation Theory we show that these two flows
are received by $M_p$ with opposite coupling constants of the
$\phi_{31}$ irrelevant perturbation. 
Some comments on the massless
S-matrices of these two flows are added.
\end{abstract}
\newpage

\section{Introduction}
Many two-dimensional massless Quantum Field Theories can be seen
as Renormalization Group (RG) flows connecting an ultraviolet (UV)
Conformal Field Theory (CFT) to an infrared (IR) one~\cite{Zam}. When
an infinite number of charges is conserved the theory is
integrable. In this case 
a powerful method of investigation consists in the
Thermodynamic Bethe Ansatz (TBA)~\cite{Al1}, that can be thought as a
set of non-linear coupled integrable equations driving the RG
evolution of various quantities along the flow. In principle the TBA
equations could be deduced from the (massive or massless) S-matrix of
the theory, but the derivation is technically easy only in the case of
{\em diagonal} S-matrices. If a kink structure connecting different
colored vacua appears, one must resort to very complicated Bethe
Ansatz techniques to diagonalize the color transfer matrix and deduce
the TBA system. In spite of this, a lot of TBA systems attached to
many integrable theories have been conjectured, and extensively
checked against UV perturbative results. In the case of massless
theories, however, also the IR limit is given by a non-trivial CFT
to which the RG flows is attracted by an irrelevant integrable
operator $\Phi(x)$. This simplified sentence means that one can write an
IR-effective action approximating the behaviour of the theory at large
scales
\eq
{\cal A} = {\cal A}_{IR} + g \int d^2 x  \Phi(x) + \mbox{h.o.t.}
\en
where $\cA_{IR}$ (formally) represents the action of the IR CFT.
The ``h.o.t.''
means higher order terms, i.e. an infinite number of 
higher dimension operators that in
principle can contribute as counterterms
each one with its independent coupling
constant. In other words, this effective theory is not
renormalizable. This makes the IR perturbation theory much more
difficult than the UV one. In these two-dimensional theories, experience
tells us that the UV perturbative series usually has a finite radius of
convergence. Instead, the IR one is at best an asymptotic series. In
spite of these difficulties we shall see in this letter that the first
few orders of the IR series can be reasonably controlled and turn out
to be of interest when compared with the
{\em exact} results coming from the integration of the TBA equations. 
This allows further checks of the validity of TBA as
well as calculation of quantities that can shed new insight on the
structure of the space of RG flows, as we shall see below.

The idea of comparing TBA results with IR perturbation theory traces
back to Al.Zamolodchikov~\cite{Al3,Al4}, but only the case of $T\bar{T}$
perturbations ($T$ being the stress-energy tensor) were developed
enough. Klassen and Melzer~\cite{spectral} went a bit further, by
exploring the simplest cases of $\phi_{31}$ IR perturbations of
minimal models. Much of the inspiration of the present work comes from
their approach. IR perturbations have been dealt with in a more
systematic way, although in a rather different problem, by
Berkovich~\cite{berk}. Also this paper contains elements that have
been illuminating for our analysis.

In this letter we make use of IR perturbation theory to compare the
TBA results of two celebrated examples of integrable massless field
theories, maybe the most studied ones:
\begin{enumerate}
\item the minimal models $M_{p+1}$ perturbed by $\phi_{13}$ that
notoriously flow to $M_p$. Following~\cite{Al4} we denote these
massless theories by $\mn_{p+1}$.
\item the $\bZ_{p-1}$-parafermionic theories perturbed by their
$\Psi_1=\psi_1\bar{\psi}_1^+\psi_1^{\dagger}\bar\psi_1^{\dagger}$
operator ($\psi_1$ being the generating parafermion), 
which also are known to flow down to $M_p$. Following~\cite{fatal} 
they will be denoted in this paper by $\prf_{p-1}$.
\end{enumerate}
That the minimal models and the parafermion theories have very strict
relation in general, even at the pure conformal level, is not a new surprise
and can be traced back to the fact that both heavily involve the
$A_1^{(1)}$ affine algebra in the deep of their constructions.
The interesting fact here is that both theories flow down to $M_p$
``attracted'' by the same operator $\phi_{31}$. One can wonder which
is the feature of the IR $\phi_{31}$ perturbation theory that
distinguishes the two flows. Our investigation has the aim of
clarifying this issue and add a new piece of information to the
beautiful puzzle of understanding the map of 2 dimensional integrable
flows. 

\section{TBA equations for $\mn_p$ and $\prf_p$ models}
A set of TBA equations for $\mn_p$ has been proposed by
Al.Zamolodchikov~\cite{Al4}, and checked against various tests that
give more than reasonable confidence to its correctness. Basically the
prediction for the scaling central charge $c(r)$ ($r=MR$ is an
adimensional quantity parametrizing the RG flow in terms of the mass
scale $M$ and inverse ``temperature'' $R$) is
\eq
c(r) = \frac{3}{\pi^2}\int_{-\infty}^{+\infty}\sum_a
d\theta \nu_a(\theta) L_a(\theta)
\label{A}
\en
where $\nu_a(\theta)=\frac{r}{2}e^\theta
\delta_{a,1}+\frac{r}{2}e^{-\theta} \delta_{a,p-2}$, the index $a$
running from 1 to $p-2$ along a $A_{p-2}$ Dynkin diagram. The
$L_a(\theta)=\log(1+e^{-\varepsilon_a(\theta)})$ are evaluated as
solutions of the set of integral equations
\eq
\nu_a(\theta)=\varepsilon_a(\theta)+\frac{1}{2\pi}\int_{-\infty}^{+\infty}
d\theta' \varphi(\theta-\theta') \sum_b l_{ab}L_b(\theta')
\label{B}
\en
with kernel $\varphi(\theta)=1/\cosh(\theta)$ and $l_{ab}$ incidence
matrix of the $A_{p-2}$ diagram. With this TBA one can test even for
low $p$ the validity of the results on the $\mn_p$ 
models established perturbatively by A.Zamolodchikov for high $p$.

A similar TBA set has been proposed by Fateev and
Al. Zamolodchikov~\cite{fatal} to describe, or even better, to give
the first evidence, of highly {\em nonperturbative} flows between
parafermions and minimal models, namely the $\prf_p$ theories briefly 
described above. Again, the TBA equations and the formula for the
scaling central charge are given by eqs.(\ref{A}) and (\ref{B}), with
the same kernel, the only difference now consisting in a different
$l_{ab}$ incidence matrix, namely that of the $D_p$ Dynkin diagram. 

\section{IR perturbation theory}
The discussion of IR perturbative series is basically plagued by the
existence of infinitely many irrelevant operators $O_j(x)$ 
that can contribute
to it each one with its independent coupling $g_j$, thus making the theory
not renormalizable. However, conformal perturbation theory has the
advantage to be strongly constrained by the fact that the operators
appearing in the correlators in the perturbative series must be found
among the operator content of the unperturbed CFT itself. The
effective perturbed theory is defined by the action
\eq
\cA = \cA_{M_p} + \int d^2x \sum_{j\in\cV} g_j O_j(x)
\en
We use the
following requirements to restrict the space of IR perturbing operators $\cV$
\begin{itemize}
\item all perturbing operators must be {\em scalars},
i.e. $\Delta_j=\bar{\Delta}_j$ in order to keep 1$+$1 Lorentz invariance
\item all perturbing operators must be irrelevant, $\Delta_j>1$
\item total derivative operators do not contribute to the action
\end{itemize}
Even after these restrictions, the space $\cV$ is still extremely
large and difficult to analyze. But there is another issue which is
the main characteristic of these perturbations: their {\em
integrability}. In other words, we know that these perturbations
describe, near IR, flows having an infinite number of conserved
currents. In particular it is known that both series of models $\mn_p$
and $\prf_p$ have for all $p$, an infinite set of local integrals of
motions (IM) of spin 1 mod 2. This does not exclude for some specific
models in these series existence of {\em other} IM's,
but at least this series of IM's must be kept conserved all along the
flow. All operators destroying even only one of these currents must be
discarded from $\cV$. This is the really powerful constraint we can
put on our series. It is indeed a fact stressed by many authors 
that only operators from
the families of $[\buno]$, $[\phi_{13}]$ and $[\phi_{31}]$ have the
nice property to preserve all the IM's of spins 1 mod 2. Therefore our
space $\cV$ can be restricted to these 3 families in the following
analysis.

If $R$ denotes the radius of the cylinder where the theory is put for
TBA analysis, 
the scaling function (normalized as to give the UV and IR central
charges in the two limits $R\to 0$ and $R\to \infty$ respectively)
is given near IR by a series expansion in the correlation
functions of the $O_j(x)$ operators on the cylinder
\eq
c(r) = c_{IR} + \sum_{n=1}^{\infty} \sum_{j_1,...,j_n\in\cV}
P_{j_1,...,j_n} g_{j_1}...g_{j_n} R^{\sum_{i=1}^n y_{j_i}}
\en
where
\eq
P_{j_1,...,j_n}=12\frac{(-1)^n}{n!}R^{2-\sum_{i=1}^n y_{j_i}} \int
\prod_{i=2}^{n} d^2u_i \langle O_{j_1}(0)
\prod_{i=2}^{n}O_{j_i}(u_i) \rangle_{cyl}
\en
The coupling constants $g_i$ carry a scale dimension
$y_i=2(1-\Delta_i)$ each, and as in the usual UV case, there will be a
relation with the mass scale $M$ of the form
\eq
g_i = \kappa_i M^{y_i}
\en
where $\kappa_i$ are numerical coefficients. Comparison with TBA data
can be done by first ordering the $\sum_{i=1}^n y_{j_i}=y_a$ in
decreasing order (all $y_i$ are negative for irrelevant
operators). Define
\eq
c_a = \sum_{J_a} P_{j_1,...,j_n} \kappa_{j_1} ... \kappa_{j_n}
\en
where the set $J_a$ is formed by all the possible choices of
$j_1,...,j_n$ such that $\sum_{i=1}^n y_{j_1}=y_a$. then the scaling
function $c(r)$ has the expansion
\eq
c(r)=c_{IR}+\sum_a c_a r^{y_a}
\label{C}
\en
that can be directly compared with the numerical integration of the TBA
equations. What in practice one has to do is to choose a truncation of
this series and try to fit it against the TBA numerical data. 

In order to explicitly compute the $P_{j_1,...,j_n}$ one has to
transform them from the cylinder to the sphere. Note that in the case
of secondary operators the transformation can be more complicate than
the usual one for primary operators considered in~\cite{KM2} to give
the ``sphere'' formulae for $P$'s. Luckily, as we shall see, for the
purposes of this paper, the only secondary operator we have to
consider is the $T\bar{T}$ secondary of the identity. The well-known
transformation of the stress-energy tensor creates a term
$\frac{c}{24}$ that allows a contribution from the 1-point function of
this operator on the cylinder. One must keep in mind this exception in
the following analysis. No secondary operator of families different
from $[\buno]$ can have a non-zero cylinder 1-point function.

In writing down the asymptotic series (\ref{C}) one has to analyze
which correlation functions really give a contribution and to which
order. Here we give a brief summary of this analysis in the case of
the three families of integrability preserving operators mentioned above.

The $[\buno]$ family contains as first (scalar) secondary the
determinant of the stress-energy
tensor $T\bar{T}$. Its
renormalization group eigenvalue $y=2(1-\Delta)$ is $-2$, therefore we
expect contribution from its 1-point function at $r^{-2}$ in the IR
series. Its 2-point function will contribute at $r^{-4}$, etc... Next
operator in the identity family is $:T^2\bar{T^2}:$ whose first
contribution appears at order $r^{-6}$.

The $\phi_{31}$ operator is the dominating one in the IR series. The
first contribution comes from its 2-point function $\langle
\phi_{31}\phi_{31} \rangle$ and, as $\Delta_{31}=1+\frac{2}{p}$, which
implies $y=-4/p$, it is at order $r^{-8/p}$. Next contribution is
given by the 3-point function at $r^{-12/p}$. This will be the crucial
point of our analysis, as we shall see. Further n-point functions of
$\phi_{31}$ appear at orders $r^{-4n/p}$. Secondaries of $\phi_{31}$
can give contributions at orders $r^{-4-8/p}$ or higher, as it appears
from analysis of the ``lowest'' case, namely $\langle \phi_{31}
\phi_{31}^{(2)}\rangle$. (Here $\phi_{31}^{(2)}$ denotes
a generic secondary of $\phi_{31}$ at level 2). 
The $\phi_{31}$ expansion is unaffected by its secondaries if we
do not consider terms $O(r^{-4})$.
Another interference that can appear is between
$\phi_{31}$ and $[\buno]$ family. The first term comes from $\langle
\phi_{31} \phi_{31} T\bar{T}\rangle$, this is at order
$r^{-2-8/p}$, and if we consider only terms of order less that this we
are also free of this problem.

Let us now turn to the contributions from the pure $[\phi_{13}]$
family. As $\phi_{13}$ is relevant, the dominant field of this family
allowed to be in $\cV$ is $L_{-2}\bar{L}_{-2}\phi_{13}$. Its 2-point
function contributes at order $r^{-6+8/(p+1)}$, in any case higher
than the limit we already imposed to eliminate interference between
$\phi_{31}$ and $T\bar{T}$. All other correlators of
secondaries of $\phi_{13}$ contribute at order higher than
this. Insertions of secondaries of the identity into such correlators
raise the order even more. Finally the interferences between
secondaries of $\phi_{13}$ and $\phi_{31}$ also happen at higher
orders, because the 2 and 3-point functions involving both $\phi_{13}$
and $\phi_{31}$ are all zero.

We conclude that, if we are content to consider contributions up to
$r^{-2-8/p}$ excluded, we can only consider the two operators $\phi_{31}$ and
$T\bar{T}$ in the perturbation expansion. Higher orders not only
become more complicate from the ``theoretical'' point of view, but a
careful numerical analysis shows that their contributions begin,
for a typical numerical sample of TBA data in the range $200<r<2000$,
to be in the region where the delicate equilibrium of asymptotical
convergence starts to break. Therefore we consider in the following
the effective action
\eq
\cA = \cA_{IR} + \int d^2x (g \phi_{31} + t T\bar{T})
\en

To finish this section, we have to mention another delicate issue that
appears in the perturbative series. When some operator is such that
for some integer $N$ it happens that $Ny=-2$, the corresponding
$N$-point function is divergent, even after analytic continuation in
$y$. To avoid this singularity, one should introduce a counterterm
that results in the appearance of a logarithmic term of the form
$r^{-2}\log r$ that must be taken into account when fitting with
numerical data. This happens in the model $M_p$ with $p$ even 
for the $\frac{p}{2}$-point function of $\phi_{31}$. Therefore, in the
$M_4$ model this problem arises already at the first term, i.e. the
2-point function of $\phi_{31}$. Similarly, in
the $M_6$ case the 3-point function (which is the crucial one in our
analysis) is plagued by this log-term. However, for higher $M_p$ with
even $p$, this interference moves to higher and higher $\phi_{31}$
correlators, and leaves the 2 and 3 point functions free of this
problem, allowing clean application of our argument.

\section{Numerical analysis}
We have performed a high precision (14 significant digits) numerical
integration of the TBA equations in the range $200<r<2000$ step
10. Stability of the results has been checked against increasing the number of
iterations and decreasing the integration step in $\theta$. The data
in this range have been fitted with a function of $r^{-1}$ at the
powers predicted by the IR perturbation analysis done in the previous
section, truncated at around $r^{-2.5}$. The reason for this choice is
explained a little below. The coefficient of the fitted
expansion thus found are reported in Tab.1. 
\begin{table}
\begin{center}
\begin{tabular}{|c|c|l|}
\hline
IR & model & $c(r)-c_{IR}$\\
\hline \hline
$M_5$ & $\mn_6$ &
$0.57582(7)r^{-\frac{8}{5}}-1.340(6)r^{-2}+0.949(5)r^{-\frac{12}{5}}$ \\
\hline
$M_5$ & $\prf_4$ &
$0.57582(2)r^{-\frac{8}{5}}-0.000(0)r^{-2}-0.946(1)r^{-\frac{12}{5}}$\\
\hline \hline
$M_6$ & $\mn_7$ &
$0.22691(2)r^{-\frac{4}{3}}+0.206(6)r^{-2}+0.3268(1)r^{-2}\log r$\\
\hline
$M_6$ & $\prf_5$ &
$0.22690(8)r^{-\frac{4}{3}}-0.207(1)r^{-2}-0.3268(8)r^{-2}\log r$\\
\hline \hline
$M_7$ & $\mn_8$ &
$0.11938(7)r^{-\frac{8}{7}}-0.841(4)r^{-\frac{12}{7}}+1.66(3)r^{-2}
-1.17(6)r^{-\frac{16}{7}}$\\
\hline
$M_7$ & $\prf_6$ &
$0.11938(9)r^{-\frac{8}{7}}+0.842(7)r^{-\frac{12}{7}}+0.00(3)r^{-2}
-1.17(9)r^{-\frac{16}{7}}$\\
\hline \hline
$M_8$ & $\mn_9$ &
$0.07203(0)r^{-1}-0.350(0)r^{-\frac{3}{2}}-0.15(4)r^{-2}-0.27(7)r^{-2}\log
r+0.5(5)r^{-\frac{5}{2}}$\\
\hline
$M_8$ & $\prf_7$ &
$0.07203(1)r^{-1}+0.350(2)r^{-\frac{3}{2}}-0.17(9)r^{-2}+0.28(1)r^{-2}\log
r-0.5(1)r^{-\frac{5}{2}}$\\
\hline \hline
$M_9$ & $\mn_{10}$ &
$0.04730(5)r^{-\frac{8}{9}}-0.194(3)r^{-\frac{4}{3}}
+0.98(0)r^{-\frac{16}{9}}-1.8(0)r^{-2}
+1.2(4)r^{-\frac{20}{9}}$\\
\hline
$M_9$ & $\prf_8$ &
$0.04730(6)r^{-\frac{8}{9}}+0.194(3)r^{-\frac{4}{3}}
+0.98(6)r^{-\frac{16}{9}}-0.0(1)r^{-2}
-1.2(5)r^{-\frac{20}{9}}$\\
\hline \hline
\end{tabular}
\caption{The numerical fits of TBA data against IR
perturbative series}
\end{center}
\end{table}
\begin{table}
\begin{center}
\begin{tabular}{|c|c|r|r|}
\hline
IR & model & value of $\kappa$ & theor. coeff. of $r^{-\frac{12}{p}}$ \\
\hline \hline
$M_5$ & $\mn_6$   & 0.048768029  & 0.946230 \\
\hline
$M_5$ & $\prf_4$  & 0.048767817  &-0.946228 \\
\hline \hline
$M_6$ & $\mn_7$   & 0.046196485  & $\infty$\\
\hline
$M_6$ & $\prf_5$  & 0.046196078  &$\infty$\\
\hline \hline
$M_7$ & $\mn_8$   & 0.042986482  &-0.842248 \\
\hline
$M_7$ & $\prf_6$  & 0.042986122  &0.842280 \\
\hline \hline
$M_8$ & $\mn_9$   & 0.039903430  &-0.350708 \\
\hline
$M_8$ & $\prf_7$  & 0.039903708  &0.350732 \\
\hline \hline
$M_9$ & $\mn_{10}$& 0.037120733  &-0.194105 \\
\hline
$M_9$ & $\prf_8$  & 0.037121126  &0.194123 \\
\hline \hline
\end{tabular}
\caption{The theoretical values of the coefficient
determined by the 3-point function of $\phi_{31}$. The values of
$\kappa$ needed for this calculation are deduced from the coefficient
of the 2-point function term.}
\end{center}
\end{table}

We started our analysis
with the $M_5$ IR theory. The $M_4$ case presents the mentioned 
anomaly to have
the first coefficient divergent, which results in a logarithmic
contribution. On the other hand, it has been extensively
discussed by Klassen and Melzer in~\cite{dseries}. They give arguments
to support the hypothesis that the two flows from $M_5$ and $Z_3$ come
into $M_4$ from opposite values of the $\phi_{31}$ coupling $g$. 
Log-terms are totally absent in the next model $M_5$. The dominating
coefficient comes from the 2-point function of $\phi_{31}$. It can be
evaluated theoretically by resorting, e.g. to formula (2.28)
of~\cite{spectral}, which is valid also for the irrelevant $\phi_{31}$
because it is a primary field. Out of that formula we extract the
coefficient $\kappa$ relating the $\phi_{13}$ coupling $g$ to the mass
scale, $|g|=\kappa M^{-4/p}$. Then use of this $\kappa$ and of the
structure constant $C_{(31),(31)}^{(31)}$ that one can get
from~\cite{dotfat} allows to compute the ``theoretical'' value of the
coefficient of the 3-point function. The only case where this does not
work is $M_6$ where the 3-point function of $\phi_{31}$ diverges
leading to appearance of a log-term and
also interferes with $T\bar{T}$. In all other cases we are able to
report the theoretical coefficient that we use as a check of the
numerical correctness of our data. The most important thing that one
immediately sees from Tab.1 is that the coefficient of the 3-point
function of $\phi_{31}$ in the $\mn_{p+1}$ model is always opposite to
that of the $\prf_{p-1}$ one, a clear sign that {\bf the two flows enter
the IR model from opposite directions}! This definitely settles
the problem of the two $\phi_{31}$ attracted flows that enter each
minimal model. But a more careful analysis of the numerical data of
Tab.1 show other interesting surprises. For example, within numerical
error it seems plausible to take, at least in all models with {\em odd}
$M_p$ as IR, the coupling with $T\bar{T}$ equal to zero for the
parafermionic flows, while it is different from zero for the minimal
models ones. 
If this fact has a relation with the different forms (with of
without $\theta$ shift -- see below) of the two massless S-matrices of
the models is an open and interesting issue.
For {\em even} $p$ instead, it happens that the
$T\bar{T}$ coupling always interferes with some $n$-point $\phi_{31}$
function. It is curious to note that the renormalized couplings still
show the desired behaviour (equal absolute value and opposite/equal
signs for odd/even $n$).

Our numerical data seem to be reliable up to some $r^{-2.5}$.
Introduction of terms higher than this do not seem to
improve the fit, in some cases even they make convergence worse. This
is a typical behaviour of asymptotic series, adding new terms can give
worse approximations of the function after a certain point. It could
be possible that this $r^{-2.5}$ plays a role of a limit after which
asymptotic convergence becomes more problematic. Luckily, this also is
the limit above which secondaries, interferences between different
families and other such phenomena are confined (at least for $p<16$), 
a lucky situation
indeed allowing us to observe the sign change in $g$ so important from
the physical point of view, before it becomes hidden by all these more
or less uncontrollable contributions. Note that number of $\phi_{31}$
correlators which fall before this limit {\em increases} with
$p$. Curiously, higher minimal models allow a better IR analysis than
the lower ones. Indeed starting from $M_7$ we are able to see also the
4 and 5-point functions. The 4-point function, as expected, gives the
same coefficient in the two flows, while the 5-point again presents
the change of sign, thus confirming once more that the two flows come to
$M_p$ from opposite directions.

\section{Some comments on the S-matrices}
For the two classes of flows under considerations, massless S-matrices
can be written. The minimal models case has been analyzed
in~\cite{FSZ2}, in relation with the so called ``imaginary
Sine-Gordon'' model of relevance for polymer physics. The $S_{LL}=S_{RR}$
matrix of $\cM_p$ is formally equal to the S-matrix of the $M_{p-1}$ model
perturbed by $\phi_{13}$ in the {\em massive} direction, and therefore
it is given by the one described in~\cite{BLc}. The $S_{RL}$ and
$S_{LR}$ blocks, instead, turn out to be proportional to the $S_{LL}$
block, calculated for a value of $\theta$ shifted by an imaginary
quantity $\alpha$.
\eq
S_{LR}(\theta)\propto S_{LL}(\theta+i\alpha) \virg
S_{RL}(\theta)\propto S_{LL}(\theta-i\alpha)
\label{undici}
\en
The shift can be fixed by requiring that the two ``mixed'' components
are equal, which should correspond to requiring time-reversal
invariance. This imposes $\alpha=\frac{n\pi}{2\mu}$
($\mu=\frac{1}{p-1}$), $n=0,1,...$. As shown in ref.~\cite{FSZ2} the
correct choice for this case is given by
$\alpha=\frac{\pi}{2\mu}$. The proportionality factor is fixed by the
analytical crossing-unitarity requirement.

For what concerns the $\prf_p$ flow, the massless S matrix can be
obtained by the following argument. Fateev~\cite{F} wrote the massive
S matrix of the $H^{(0)}_p$ theories, i.e. the $\bZ_p$-parafermionic
models perturbed by their fundamental parafermion operator in the {\em
massive} regime. He got the S matrix as quantum group reduction of the
S matrix of the {\em Sausage model} (see ref.~{FOZ}). This latter is
basically the spectral parameter dependent R-matrix of the $sl(2)_q$
quantum group in the spin 1 representation (in contrast to the
Sine-Gordon S matrix, which is in the spin $\half$ representation). At
$q$ root of 1, the quantum group reduction gives the S matrix of the
$H^{(0)}_p$ models, as one can easily check against the known cases
$p=2,3,4$. If this procedure works fine in the massive sector, we do
not see any reason preventing its extension to the massless
regime. The S-matrix for the massless Sausage model is written
in~\cite{FOZ}, and consists of blocks all equal to the Sine-Gordon
S-matrix, i.e. $S_{LL}=S_{RR}=S_{RL}=S_{LR}=S_{SG}$. This is not
unexpected, if we remember that the IR limit of the massless sausage
is the $c=1$ CFT, that is also the UV limit of Sine-Gordon theory. The
$q$-reductions of this S-matrix describe the parafermionic flows
$\prf_p$. This also implies that the $S_{LL}$ and $S_{RR}$ components
are equal for the two flows $\mn_{p+1}$ and $\prf_{p-1}$. Again, this
is not surprising, as the two flows share the same IR limit $M_p$. The
distinction between the two cases is given by the $S_{LR}=S_{RL}$
blocks that in the $\mn_{p+1}$ case have the shift in $\theta$,
eq.(\ref{undici}) while in the $\prf_{p-1}$ case
$S_{LR}(\theta)=S_{LL}(\theta)$.

This observation would exhaust the problem of writing the massless
S-matrices for the two series of flows examined in this
paper. However, one has to check the coincidences $\prf_2$=$\mn_3$
(Ising model) and $\prf_3$=$\mn_5$ (3 state Potts Model). While the
first one does not give any problem, the Potts case is rather
intriguing. The $S_{LL}$ and $S_{RR}$ blocks coincide, as expected,
but the $S_{LR}=S_{RL}$ do not. On the other hand, a TBA analysis
similar to the one done in~\cite{FSZ2} (and generalizing the massive
case of~\cite{Al3} for the scaling tricritical Ising model) shows that
both S matrices give the correct $c_{UV}=\frac{4}{5}$ and
$c_{IR}=\frac{7}{10}$ (note that that the S matrix
with unshifted $S_{LR}$ giving $c_{UV}=\frac{9}{5}$ in~\cite{FSZ2}
differs from ours by a non-trivial Z-factor). It is evident that there
must be a transformation of the asymptotic particle basis relating one
S matrix to the other. The most appealing feature in this respect is
the fact that the two theories $\mn_5$ and $\prf_3$ are in a sense
related one another by an orbifold procedure. Consider their UV
limits: the $\mn_5$ is the {\em diagonal} minimal model $M_5$, the
$\prf_3$ one is the $\bZ_3$-parafermion, which coincides with the
{\em complementary} modular invariant ($(A_4,D_4)$ in the Cappelli,
Itzykson and Zuber classification~\cite{CIZ}). One passes from one to
the other modular invariant by an orbifold procedure. If we make the
hypothesis that also off-criticality the two models are related one
another by such an orbifold, the S matrix of the $\prf_3$ model will
be obtained from the $\mn_5$ one by the orbifold recipe explained
in~\cite{FG}. This indeed works fine in the massive analogs of these
two models, as well as for the $S_{LL}=S_{RR}$ blocks here, which turn
out to be equal in the two cases, thanks to the automorphism $A_3
\equiv D_3$ of their adjacency diagrams. However, straightforward
application of this folding procedure to the $S_{LR}=S_{RL}$ blocks
would predict also for them equality in the two models, while, as
seen, they are definitely {\em not} equal in this case. Our opinion is
that the folding procedure \`a la Fendley Ginsparg~\cite{FG} has to be
modified slightly in the massless case, in the sense that the role of
left and right movers are inverted: if we denote by $-,0,+$ the 3
nodes of $A_3$, the $R_+$ right mover behaves exactly as $L_-$ left
mover in the folding procedure, and so on. The $- \leftrightarrow +$
symmetry of $A_3$ guarantees that if only right or only left objects
are concerned ($S_{LL}$ and $S_{RR}$) this inversion is uninfluent. If
instead we are considering the scattering of a left and a right movers
this modifies the folding procedure in such a way to give the shifted
$S_{LR}$ from the unshifted one (they are related one another by
sending $\sinh \to i \cosh$ in this case. We intend to return to this
curiosity and to the general problem of writing S matrices for
non-diagonal $\phi_{13}$ perturbed minimal models in massive and
massless regimes in future.

\section{Conclusions}
In this letter we have given evidence that the two flows entering any
minimal model from $\phi_{31}$, namely the integrable theories
$\mn_{p+1}$ and $\prf_{p-1}$, come into their IR limit from opposite
directions. This result can be interesting in itself, as it sheds
further light on the map of the two-dimensional RG space of integrable
flows and of minimal models in particular. Further things should be
investigated to better understand this issue. We have seen in the
previous section that the ``fine structure'' distinction between
different modular invariants may play a role in solving apparent
contradictions between S matrices that appear to be different and are
expected to be equal, as in the commented $\prf_3$ versus $\mn_5$
case. Also, the flows examined in this paper are not unique: the
parafermions themselves have in general more than one modular
invariant. It would be interesting to solve the whole puzzle of the
flows between all parafermions and minimal models ones. This could
also point out, by ``undoing the truncation'' procedures~\cite{FSZ2},
to some orbifold version of the imaginary Sine-Gordon and of the
Sausage models. Work in this direction is in progress.

We think that the most appealing fact of the result of the present
paper is the
way it has been obtained, i.e. by use of IR perturbation, 
an instrument used only marginally so far in TBA analysis. It is
possible that a better development of this IR techniques, maybe by
comparison with the similar ones developed in~\cite{berk}, could provide
new tools of investigation of the massless integrable theories.
\vskip 0.3cm

{\bf Acknowledgments} - We thank A.Berkovich, P.Dorey, R.Tateo and K.Thompson
for useful discussions. The last three are also acknowledged for
having pointed out a minor misprint in Tab. 1 in the first preprint version
of this paper.
F.R. is grateful to ENSLAPP-Annecy for the kind and warm hospitality
extended to him during part of this work.
This work is supported in part by NATO Grant CRG 950751.

\end{document}